\documentclass[preprint]{revtex4}
\usepackage{amssymb}
\usepackage{amsmath}
\usepackage{graphicx}

\begin{document}

\title{Dark energy from gravitational collapse?}
\author{L\'{a}szl\'{o} \'{A}. Gergely}
\affiliation{Departments of Theoretical and Experimental Physics, University of Szeged,
Hungary}

\begin{abstract}
We discuss the status of both cosmological and black hole type singularities
in the framework of the brane-world model of gravity. We point out that the
Big Bang is not properly understood yet. We also show new features of the
gravitational collapse on the brane, the most important being the production
of dark energy during the collapse.
\end{abstract}

\email{gergely@physx.u-szeged.hu}
\date{\today }
\startpage{1}
\endpage{}
\maketitle

Singularities, both cosmological and of black hole type, are intriguing
objects predicted by general relativity. One of the most desired aims of
quantum gravity is to avoid such infinities. Although interesting results
stem out both from loop quantum gravity and string theory, no conclusive
decision has been reached. In the absence of a final theory of quantum
gravity the question arises how the recently introduced, modified, but still
classical theories of gravity, motivated by string / M-theory, will tackle
these topics.

One of the most promising such theory is the generalized Randall-Sundrum
second model, introduced in its original form in \cite{RS2}, in which our
observable universe is a 4-dimensional hypersurface (the brane), embedded
into a 5-dimensional curved space-time (the bulk). Standard model fields
live on the brane, which is held together by a tension $\lambda $. A
positive brane tension allows to recover the Einstein-gravity on the brane,
at low energies and insignificant bulk curvature.

When this scenario is employed as a cosmological model, the brane becomes a
4-dimensional Friedmann-Lema\^{\i}tre-Robertson-Walker (FLRW) universe.
Cosmological evolution is consequence of the brane moving in the
Schwarzschild - anti de Sitter bulk. The scale factor, related in a simple
way to the distance from the bulk black hole, evolves driven by a \textit{%
modified }Friedmann and a \textit{modified} Raychaudhuri equation \cite%
{Decomp}. The modifications include a quadratic source term in the brane
energy momentum (modifying early cosmology), a dark radiation term scaled by
the mass of the bulk black hole (representing the electric part of the bulk
Weyl curvature), source terms from a possible asymmetric embedding (which
can cause late time acceleration) and the pull-back to the brane of any
existing bulk energy momentum.

The Big Bang, occurring when the scale factor is zero, corresponds to the
brane emerging from the singularity. However it would contradict basic
principles of physics to allow the brane escaping from below the horizon. In
consequence it is not clear, whether the brane somehow condensates outside
the event horizon from (quantum) bulk fields, or the bulk black hole is
created far away from the pre-existing brane. The latter possibility was
investigated in \cite{ChKN}, assuming a brane which is radiating into the
fifth dimension. If so, the static character of the bulk is obstructed and
its simplest model would be the Vaidya-anti de Sitter space-time.
Cosmological evolution in this radiation-filled bulk was further exploited
in \cite{Decomp}-\cite{GKK}. The parameters of the model can be fitted in
such a way that the late time universe is extremely close to what is
predicted by general relativity and verified by astronomical observations.

Black holes are fascinating objects, first predicted by general relativity.
The energy momentum, as source of gravity, curves space-time in the vicinity
of these objects to such an extent, that even light rays can follow circular
orbits, like the planets revolve about stars. Strangely enough, much closer
to the center, the role of time and space is interchanged in some subtle
way. The curvature of space-time approaches infinity near the center,
forming a space-time singularity. According to the Cosmic Censorship
hypothesis, an event horizon luckily hides the singularity from any outside
viewer in the majority of cases. Even so, black holes are dangerous
creatures. They accrete surrounding matter in galactic centers and force the
magnetic field lines into configurations, which give birth to powerful jets.
These are starring on some of the most astonishing astronomical pictures.

It is not surprising, that the search for black holes in the brane-world
scenario has consumed much energy. But we face a series of negative results.
In brane cosmological models there are black holes in the bulk, but not on
the brane, where they are allowed just as minuscule test particles. Looking
for spherically symmetric vacuum solutions on the brane, the Schwarzschild
black hole turns out as the unique solution of the modified gravitational
equations, provided (a) the bulk is empty (b) the Weyl curvature of the bulk
has no electric part, and (c) the embedding of the brane is symmetric.
However the singularity has to extend into the fifth dimension \cite{ChRH}.
The emerging black string could decay into a black cigar, due to the
Gregory-Laflamme instability \cite{GL}. However, more recently it was shown
that under very mild assumptions, classical event horizons cannot pinch off 
\cite{HorowitzMaeda}. 

Becoming a black string, the black hole singularity, resulted from the
gravitational collapse of a stellar object on the brane leaves its cradle.
Such a model, with black strings puncturing the cosmological brane was
investigated in \cite{SwissCheese}. The brane shows a Swiss-cheese type
structure, with Schwarzschild black \textquotedblright
holes\textquotedblright\ in a \textquotedblright cheese\textquotedblright\
represented by the cosmological perfect fluid. In certain cases such a brane
evolves towards a new type of \textit{pressure singularity}, accompanied by
a \textit{regular} behavior of the cosmic acceleration.

By turning on the electric part of the bulk Weyl curvature, the modified
Einstein equations are solved by the Reissner-Nordstr\"{o}m black hole with
mass $m$ and tidal charge $q$ \cite{tidalRN}. The tidal charge arises from
the Kaluza-Klein modes of gravity, as opposed to general relativity, where $%
q=Q^{2}$ represents the square of the electric charge $Q$. \ When $q>0$, the
tidal black hole is analogous to the charged black hole: it has two
horizons, both below the Schwarzschild radius. A negative tidal charge, by
contrast, increases the radius of the event horizon. Gravity is strengthened
on the brane, light deflection and gravitational lensing are stronger, than
for a Schwarzschild black hole with the same mass. The bulk containing such
a tidal black hole brane remains unknown.

Gravitational collapse of a dust sphere in general relativity results in a
black hole. Does this phenomenon, known as the Oppenheimer-Snyder collapse 
\cite{OppSny} generalize to brane-worlds? The question was answered in \cite%
{BGM}. The collapse of a homogeneous Kaluza-Klein energy density with static
exterior leads to either a bounce, black hole formation or to a naked
singularity. In the absence of the tidal charge the same analysis shows that
the vacuum surrounding the collapsing sphere cannot be static.

This no-go result is based on the assumption that the collapsing matter is
dust, but this can obviously be relaxed. We do this for the case of
vanishing Kaluza-Klein modes. In other words, we consider a collapsing FLRW
sphere with constant comoving radius $\chi _{0}$ (and physical radius $%
R=\chi _{0}a$), embedded into a Schwarzschild brane \cite{collapse}.
Switching out the tidal charge has better motivation as the desire for
simplicity. Whenever the radius of the collapsing sphere is much higher than
the characteristic scale of the extra dimension, the black-hole metric on
the brane is Schwarzschild \cite{EHM}. Whereas table-top experiments \cite%
{tabletop} on possible deviations from Newton's law currently probe gravity
at submillimeter scales, this seems no severe restriction.

With a collapsing star composed of perfect fluid with energy density $\rho $
and pressure $p$, the vacuum exterior with Schwarzschild mass $m$ can be
held static. For vanishing brane cosmological constant and flat spatial
sections the generic junction conditions derived in \cite{NoSwissCheese}
imply the following evolution of the scale factor $\ a$ in comoving time $%
\tau $:%
\begin{equation}
a^{3/2}=a_{0}^{3/2}-\left( \frac{9m}{2\chi _{0}^{3}}\right) ^{1/2}\tau \ .
\label{atau}
\end{equation}%
Here $a_{0}$ represents the initial value of the scale factor. Inevitably,
after comoving time $\tau _{1}=\left( 2\chi _{0}^{3}a_{0}^{3}/9m\right)
^{1/2}$, the collapse ends in the singularity $a=0$.

The Schwarzschild mass is related to the \textquotedblright
physical\textquotedblright\ mass $M=(4\pi R^{3}/3)\rho $ as%
\begin{equation}
m=M\left( 1+\frac{\rho }{2\lambda }\right) \ .  \label{mM}
\end{equation}%
How can $m$ be interpreted from the interior, if different from the
\textquotedblright physical\textquotedblright\ mass? As shown in \cite%
{collapse}, $m$ represents the Bardeen quasilocal mass \cite{Bardeen},
summing up both matter and gravitational energy contributions. Both masses
cannot be constant, as they differ by a typical brane-world contribution $%
\rho /\lambda $ (due to the source term quadratic in the energy momentum).

The Schwarzschild mass being held constant, $M$ varies in a precise way
during the collapse. The ensemble of the junction conditions and modified
Einstein equations show that both $\rho $ and $p$ increase in magnitude
towards infinite values \cite{collapse}. Eq. (\ref{mM}) then implies that $M$
should accordingly decrease, until it reaches zero. Where has the
\textquotedblright physical\textquotedblright\ mass disappeared? We obtain
the answer from the equation of state of the fluid:%
\begin{equation}
\frac{p}{\lambda }=\frac{1}{2}\left( 1-\frac{\rho }{\lambda }\right) -\frac{1%
}{2}\left( 1+\frac{\rho }{\lambda }\right) ^{-1}\ .  \label{EOS}
\end{equation}%
At the beginning of the collapse (when $\rho \ll \lambda $ holds) this
approaches 
\begin{equation}
p\thickapprox -\frac{\rho ^{2}}{2\lambda }\ .  \label{EOS0}
\end{equation}%
This represents ordinary matter, satisfying all energy conditions. It is
basically dust (a tiny negative pressure appears in the first order).
Towards the final stages of collapse, when $\rho \gg \lambda $ Eq. (\ref{EOS}%
) gives%
\begin{equation}
p\thickapprox -\frac{\rho }{2}\ .  \label{EOSfinal}
\end{equation}%
The magnitude of the negative pressure has considerably increased such that
the condition for dark energy $\rho +3p\thickapprox -\rho /2<0$ is
satisfied. Thus \textit{in the brane-world scenario the original matter
configuration is turned into dark energy by gravitational collapse}.

The high negative pressure (tension) has the same effect as a positive
cosmological constant. Otherwise stated, a tension is raising in the brane,
slowing down gravitational contraction. In vain however, the contraction
goes on until the singularity is formed. This is due to the quadratic source
term, which dominates at high energy densities.

Note that in the general relativistic limit the tension vanishes, the fluid
turns into dust and we recover the Oppenheimer-Snyder collapse.

We conclude with a few remarks. First, the notorious difficulty in finding
black hole solutions on the brane may be related to the tensions arising in
the gravitational collapse. Second, when black holes are formed on the
brane, the initial configuration of ordinary matter surprisingly turns into
dark energy. However, in brane-worlds even dark energy speeds up
gravitational collapse. Third, a similar, but higher dimensional mechanism
to what we have just described here may be responsible for the initial
condensation of branes (containing dark energy) from a primordial bulk. This
would solve the Big Bang problem.

\end{document}